\pdfoutput=1

\documentclass[twocolumn,showpacs,amssymb,aps,nofootinbib,floatfix,superscriptaddress]{revtex4-1}

\usepackage{epsfig}
\usepackage{hyperref}

\newcommand{\br}[1]{\langle #1\rangle}

\begin{document} 

\title{\bf A simple model for rapidity fluctuations in the initial state of ultra-relativistic heavy-ion collisions}

\author{Wojciech Broniowski}
\email{Wojciech.Broniowski@ifj.edu.pl}
\affiliation{The H. Niewodnicza\'nski Institute of Nuclear Physics, Polish Academy of Sciences, 31-342 Cracow, Poland}
\affiliation{Institute of Physics, Jan Kochanowski University, 25-406 Kielce, Poland}

\author{Piotr Bo\.zek}
\email{Piotr.Bozek@fis.agh.edu.pl}
\affiliation{AGH University of Science and Technology, Faculty of Physics and
Applied Computer Science, al. Mickiewicza 30, 30-059 Cracow, Poland}

\begin{abstract}
Two-particle pseudorapidity correlations are analyzed in a simple model, where in the initial stage 
of the reaction multiple sources, extended in rapidity, are created. We show how the fluctuations
of the length of the sources in rapidity generate correlations in the initial entropy deposition, which 
later contribute to the observed longitudinal correlations in hadron production. Our analysis, which is 
analytic and leads to straightforward formulas, allows us to understand the structure of the correlations, 
in particular to identify the component related to the fluctuation of the numbers of sources and the component from the length fluctuations.
We also present the results in terms of the expansion in the basis of the Legendre polynomials.
A number of further effects is discussed, such as smearing of the pseudorapidity distributions or resonance decays. 
Our results reproduce qualitatively and semiquantitatively the basic features of the recent measurements at the LHC. 
\end{abstract}

\date{30 November 2015}



\maketitle

\section{Introduction}

Recently, new data for the pseudorapidity correlations in Pb-Pb, p-Pb, and p-p collisions at the Large Hadron Collider energies 
were released by the ATLAS Collaboration~\cite{ATLAS:2015kla,ATLAS:anm}. These studies continue the longstanding 
experimental~\cite{Uhlig:1977dc,*Alpgard:1983xp,*Ansorge:1988fg,*Alexopoulos:1995ft,*Back:2006id,*Abelev:2009ag,*Adam:2015mya,*ATLAS:2012as,*De:2013bta} 
and theoretical~\cite{Bzdak:2012tp,Dusling:2009ni,*Bzdak:2009xq,*Amelin:1994mf,*Braun:1997ch,*Yan:2010et,*Brogueira:2006yk,*Bialas:2011xk,%
*Olszewski:2013qwa,*Olszewski:2015xba,Bhalerao:2014mua,Jia:2015jga,Monnai:2015sca} efforts aimed at understanding of this important phenomenon, 
sensitive to the dynamics of the collision in its earliest stages. 

\begin{figure}
\vspace{-10mm}
\begin{center}
\includegraphics[width=0.35 \textwidth,angle=-90]{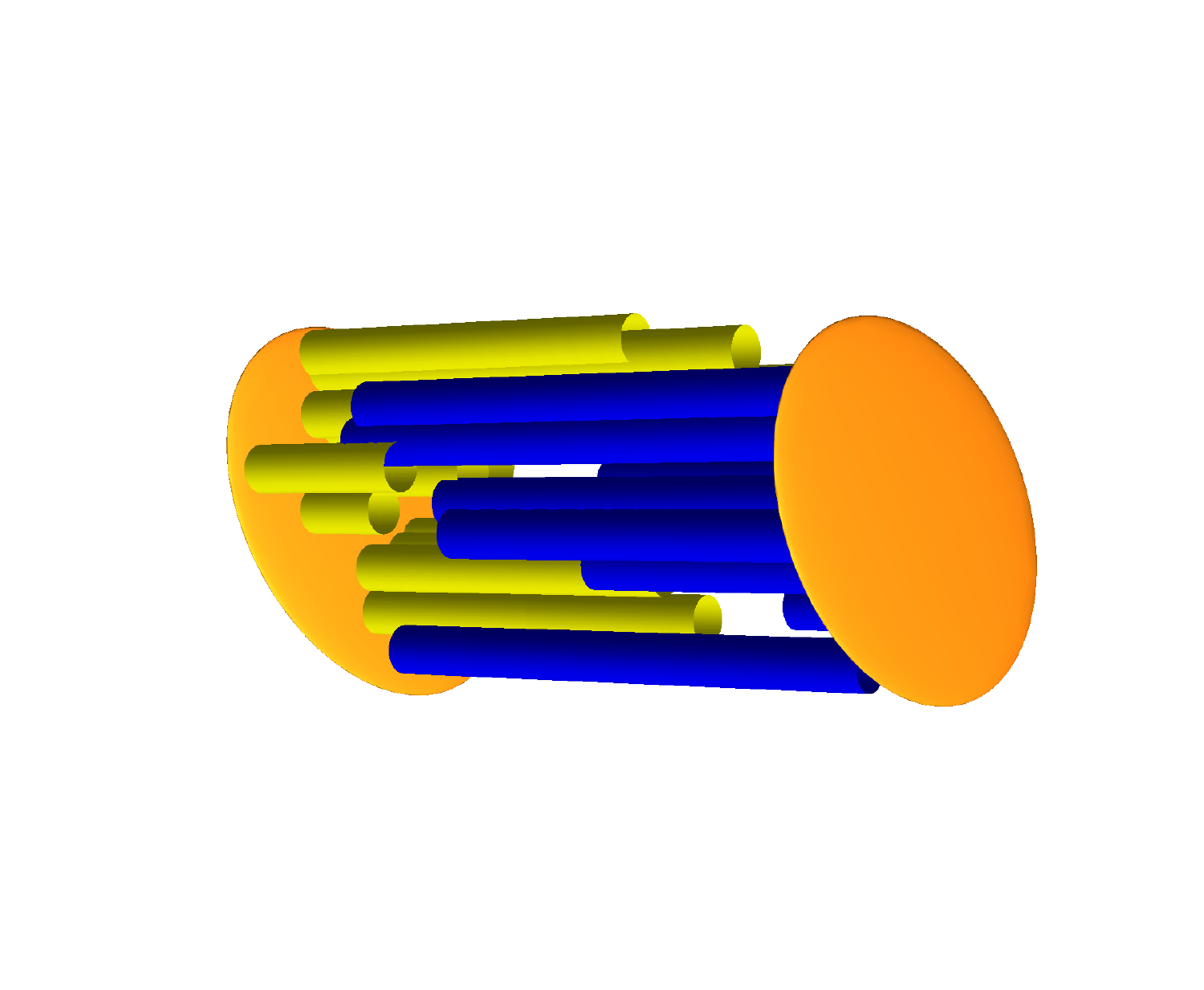} 
\end{center}
\vspace{-17mm}
\caption{Sources of fluctuating length, extending  along the spatial pseudorapidity. The cartoon shows the situation in the early stage, right 
after the collision.  \label{fig:tubes}}
\end{figure}

This paper is devoted to theoretical understanding of the newest results~\cite{ATLAS:2015kla,ATLAS:anm}.
Our study is based on what we call the {\em longitudinally-extended source model}. The initial entropy distribution 
in the longitudinal direction originates from the decay of strings or flux tubes connected to excited charges in the two colliding nuclei. The charges can be 
thought of as the wounded nucleons~\cite{Bialas:1976ed,*Bialas:2008zza} or wounded quarks~\cite{Bialas:1977en,*Anisovich:1977av}, thus are associated 
(attached) to a given colliding nucleus (cf. Fig.~\ref{fig:tubes}). A crucial aspect of the model is that the longitudinal position of the other end-point of the source 
is random, uniformly distributed in the central rapidity range. The use
of the uniform distribution is a simplification, justified for the relatively narrow interval where the correlations 
are presently measured (pseudorapidity range $[-2.4,2.4]$ in the ATLAS experiment~\cite{ATLAS:2015kla,ATLAS:anm}). 
Moreover, it allows for a straightforward analytic evaluation of the two-particle correlation function. 
This simple model grasps the essential features of realistic  
Monte Carlo models implementing the QCD string decays in the first stage of the collisions~\cite{Andersson:1983ia,*Wang:1991hta,*Lin:2004en}.
The idea that entropy deposition originates from string-like objects whose other end-point is randomly
distributed in pseudorapidity $\eta$ is related to the Brodsky-Gugnon-Kuhn mechanism~\cite{Brodsky:1977de}). It has also been discussed in the description 
of the fragmentation region in high energy collisions~\cite{Bialas:2004kt}.

The study of multiplicity correlations can be conveniently done by projecting the 
two-particle correlations function on a two-dimensional basis of orthogonal
polynomials~\cite{Bzdak:2012tp}. This method has been applied to p-p, p-Pb,  and Pb-Pb collisions by the
ATLAS Collaboration~\cite{ATLAS:anm} and was the subject of theoretical studies~\cite{Bzdak:2012tp,Jia:2015jga,Bozek:2015tca,Bzdak:2015dja,Monnai:2015sca,Pang:2015zrq}.

The analytic formulas obtained for the two-particle correlation functions allow us to understand the structure of the correlation function and to
identify various components. As noticed in Ref.~\cite{Bzdak:2012tp}, the fluctuations of the numbers of sources alone generate non-trivial 
longitudinal correlations. We show, however, that the dominant effect comes from the correlation due to the emission of 
two particles from the same source, which is caused by the length fluctuations. This effect leads to the pair emission 
probability with a relatively long range in rapidity.

We find generically that the magnitude of the coefficients of the projection of the two-particle correlation functions on orthogonal polynomials is inversely proportional  to the number of sources. 
With a natural proportionality relation between the number of sources and the final multiplicity, it can explain the experimentally observed scaling of these coefficients 
with the inverse number of observed hadrons, and the universality in A-A, p-A, and p-p collisions~\cite{ATLAS:anm}. 

We recall that the presence of the long-range fluctuations of the entropy density 
in the longitudinal direction is also visible in the experimental data
for the event-plane decorrelation in pseudorapidity~\cite{Khachatryan:2015oea}.
The mechanism of the flow angle decorrelation has been discussed in a number of papers%
~\cite{Bozek:2010vz,Petersen:2011fp,Xiao:2012uw,Pang:2014pxa,Jia:2014ysa,Bozek:2015bha,Bozek:2015bna}.
Fluctuations in the entropy deposition at different space-time rapidities lead to 
a relative torque of the elliptic and triangular flow angles at forward and backward rapidities.. 
The fluctuating end-point mechanism discussed here has been applied numerically in Ref.~\cite{Bozek:2015bna} 
to describe the event-plane decorrelation (the torque effect~\cite{Bozek:2010vz}) reported by the CMS Collaboration~\cite{Khachatryan:2015oea},
where it was essential to describe the p-Pb data. A conceptually similar model for the formation of the initial state for the hydrodynamic evolution, based on 
wounded quarks, was introduced in Ref.~\cite{Monnai:2015sca}.

\section{Experimental measures \label{sec:measure}}

The preliminary experimental results of Ref.~\cite{ATLAS:2015kla,ATLAS:anm} refer to correlations of two different charged hadrons of pseudorapidities 
$\eta_1$ and $\eta_2$, measured in a given centrality class. The relevant quantity is defined as 
\begin{eqnarray}
C(\eta_1, \eta_2) &=& \frac{ \br{\rho(\eta_1,\eta_2)} }{ \br{\rho(\eta_1)}  \br{\rho(\eta_2)}}= \frac{S(\eta_1,\eta_2)}{B(\eta_1,\eta_2)}, \label{eq:C}
\end{eqnarray}
where $\rho(\eta_1,\eta_2)$ and $\rho(\eta_{1,2})$ are the distributions of pairs and single particles, with $\br{.}$ denoting the averages over events. 
In the experiment, the quantity is estimated by taking the ratio of the {\em signal} histogram $S$ with physical pairs over the histogram $B$ formed of 
mixed-event pairs. 
The ATLAS Collaboration~\cite{ATLAS:2015kla}  uses a measure obtained from $C(\eta_1, \eta_2)$ by dividing it by its marginal distributions, namely
\begin{eqnarray}
C_N(\eta_1, \eta_2) = \frac{C(\eta_1, \eta_2) }{C_p(\eta_1)C_p(\eta_2)}, \label{eq:CN}
\end{eqnarray}
with
\begin{eqnarray}
C_p(\eta_1)=\int_{-Y}^Y \!\!\!\!\!\! d\eta_2 \, C(\eta_1, \eta_2) ,  \;\;\; C_p(\eta_2)=  \int_{-Y}^Y \!\!\!\!\!\! d\eta_1 \,C(\eta_1, \eta_2), \nonumber \\ \label{eq:cp}
\end{eqnarray} 
where $[-Y,Y]$ is the acceptance range for pseudorapidities $\eta_{1,2}$. In the ATLAS setup $Y=2.4$. This transformation reduces the effects of the
overall multiplicity fluctuations on the shape of the correlation function.
In addition, the correlation functions are conventionally normalized to 1, i.e., one introduces
\begin{eqnarray}
 \overline{C}(\eta_1, \eta_2)  =  \frac{C(\eta_1, \eta_2)}{\int_{-Y}^Y \! d\eta_1 \int_{-Y}^Y \! d\eta_2 \,C(\eta_1, \eta_2) } \label{eq:Cbar}
\end{eqnarray}
and similarly  $\overline{C}_N(\eta_1,\eta_2)$.

The $a_{nm}$ coefficients are defined via the expansion of the correlation functions in a basis of orthonormal functions~\cite{Bzdak:2012tp}:
\begin{eqnarray}
a_{nm} &=& \int_{-Y}^Y \frac{d \eta_1}{Y} \int_{-Y}^Y \frac{d \eta_2}{Y} \overline{C}(\eta_1,\eta_2) T_n\left(\frac{\eta_1}{Y}\right) T_m\left(\frac{\eta_1}{Y}\right), \nonumber \\
a_{nm}^N &=& \int_{-Y}^Y \frac{d \eta_1}{Y} \int_{-Y}^Y \frac{d \eta_2}{Y}  \overline{C}_N(\eta_1,\eta_2) T_n\left(\frac{\eta_1}{Y}\right) T_m\left(\frac{\eta_1}{Y}\right). \nonumber \\
\end{eqnarray}
The choice made in Ref.~\cite{ATLAS:2015kla,ATLAS:anm,Jia:2015jga} is 
\begin{eqnarray}
T_n(x)&=&\sqrt{\frac{2n+1}{2}}P_n(x), 
\end{eqnarray}
where $P_n(x)$ are the Legendre polynomials. The normalization is such that the functions $T_n(x)$  satisfy the orthonormality condition 
$\int_{-1}^1 dx T_n(x) T_m(x)=\delta_{nm}$.

\section{Correlation function in the longitudinally-extended source model \label{sec:flufun}}

In a wide class of approaches,  the initial stage of the  ultra-relativistic $A+B$ reaction comprises  individual collisions 
between nucleons or their constituents. As a result, local deposition of entropy takes place. 
In our analysis we assume that these {\em sources} of entropy may be associated with the $A$ or $B$ nuclei. 
The prototype is the wounded nucleon model~\cite{Czyz:1969jg,Bialas:1976ed,Miller:2007ri}, but one may also think of 
wounded constituent quarks~\cite{Bialas:1977en,Anisovich:1977av,Bialas:1977xp,Bialas:1978ze} 
or diquarks~\cite{Bialas:2006kw,*Bialas:2007eg}. 
The approach was successful in describing the transverse dynamics, were the distribution of the
initial sources in the transverse plane at mid-rapidity was relevant~\cite{Miller:2003kd}. 

For the multiplicity distributions, the considered approach can be 
applied both to elementary collisions, where no substantial final
state interaction occurs, and to nuclear collisions, where particle
production happens after a collective expansion stage. 
In the former case, the source distribution in rapidity  are directly 
related to probabilities of particle production at a given rapidity,
with a possible shift in rapidity at hadronization. 
In the latter case, the  rapidity distribution from elementary sources 
forms the initial condition for the subsequent dynamical evolution of the system.
The space-time dynamics of high-energy collisions is expected to follow the inside-outside 
cascade~\cite{Bjorken:1976mk}, with particle production occurring
around a common longitudinal proper time $\sqrt{t^2-z^2}$. Particles 
with rapidity $y$ are produced approximately at the space-time rapidity 
$\frac{1}{2}\log\left(\frac{t+z}{t-z}\right) \simeq y$. If an intermediate 
collective evolution of matter occurs, a similar argument applies for 
the initial conditions of the fluid, where the Bjorken scaling flow may be assumed.
Subsequent longitudinal and transverse evolution of the medium influences 
strongly the transverse momentum spectra. The longitudinal dynamics is less 
pronounced,  especially if the longitudinal pressure is reduced due 
to non-equilibrium corrections~\cite{Ryblewski:2010bs,Martinez:2010sc}.

In the following, we study the initial conditions in space-time rapidity, assuming
that they are close to the final rapidity distribution of matter just before 
hadronization (or thermal hadron emission from the fluid). Moreover, we assume that the observed 
pseudorapidity distributions are close to the rapidity distribution of 
the emitting sources. The effect of smearing of the distribution in the process of hadronization or thermal hadron emission, 
followed with resonance decays, is estimated explicitly and found to reduce the correlations. 

For the present goal of describing the longitudinal correlations, one needs sources that extend along the space-time rapidity. 
One may think here of strings or flux-tubes, which are pulled by a nucleon or constituent from nucleus $A$ or $B$.
The model is depicted in Fig.~\ref{fig:tubes}, where the longitudinally-extended sources have fluctuating length. Our approach includes the following  components:
\begin{enumerate}
 \item Each source may be associated with a nucleon (or its constituent) belonging to nucleus $A$ or $B$.
 In a given collision we have $N_A$ sources associated to $A$ and $N_B$ sources associated to $B$. 
 We introduce the short-hand notation 
\begin{eqnarray}
N_+=N_A+N_B, \;\;\; N_-=N_A-N_B
\end{eqnarray}
for the total number of sources and their difference between $A$ and $B$.
\item The entropy deposition (particle emission) from different sources is independent from one another.
\item In the rapidity range between the end-points, the entropy deposition from a given source is uniformly distributed in rapidity.
\item The end-point of the longitudinally-extended source is randomly fluctuating, with the distribution adjusted in such a way that 
  the phenomenological~\cite{Bialas:2004su} average  distribution from a source is reproduced. In the mid-rapidity region 
  it leads to a uniform distribution of the end-point in space-time rapidity (cf. Appendix~\ref{sec:ends}).
 \item In addition, we may overlay a distribution of strength over the sources~\cite{Bozek:2013uha}, described by the random variable $\omega$.
 A statistical measure that appears in our formulas is
 \begin{eqnarray}
 s(\omega)=\frac{{\rm var}(\omega)}{\br{\omega}^2}.
 \end{eqnarray}
 \item As mentioned before, the role of a possible  intermediate evolution stage (hydrodynamics, transport) is solely to provide an event-by-event 
 mapping between the original distribution of entropy in spatial rapidity and the final distribution of hadrons in pseudorapidity. 
 \item In the late stage, we also discuss the effects of hadronization or thermal hadron emission and resonance decays.
\end{enumerate}

The details of the derivation of the correlation function $C(\eta_1,\eta_2)$ are given in Appendices \ref{sec:lesm}-\ref{sec:strength}. Since we compare the results for the model with and without the length fluctuations, 
we introduce a parameter $r$ into our formulas, with $r=1$ and $r=0$ corresponding to present or absent length fluctuations, respectively. We also introduce the short-hand notation 
\begin{equation}
u_{1,2}=\eta_{1,2}/y_b, \label{eq:u}
\end{equation}
with  
$y_b$ denoting a parameter close to the rapidity of the beam (cf. Appendix~\ref{sec:ends}). We have the following analytic expression for the correlation 
function (\ref{eq:C}) in the mid-rapidity region:
\begin{widetext}
\begin{eqnarray}
C(\eta_1, \eta_2) = 1 &+&  \frac{1}{\left[ \br{N_+} + \br{N_-}u_1 \right] \left[ \br{N_+}+ \br{N_-}u_2 \right]} \nonumber \\ &\times&
                          \Big\{ \br{N_+} \left [ r (1 - {u_1 u_2} - |u_1-u_2|) +s(\omega)(1+r + (1-r)u_1 u_2  - r |u_1-u_2| ) \right ] + \nonumber \\ &+& \br{N_-} s(\omega) (u_1+u_2) 
                            + {\rm var}(N_+)  + {\rm var}(N_-) u_1 u_2 
                            + {\rm cov}(N_+,N_-)(u_1+u_2) \Big\}, \label{eq:gen2}
\end{eqnarray}
\end{widetext}
where the statistical moments are evaluated over the events. We note the obvious symmetry $\eta_1 \leftrightarrow \eta_2$ (particles are not distinguishable). 

The correlations originate from three kinds of effects: 
\begin{enumerate}
 \item the event-by-event fluctuation of the number of sources (the last three terms in curly brackets),
 \item the overlaid fluctuation of strength of sources (the terms including~$s(\omega)$,
 \item and the length fluctuations (entering when $r=1$).
\end{enumerate}
The following structures appear: $u_1 u_2$, $u_1+u_2$ (only for asymmetric collisions $A\neq B$), and $|u_1-u_2|$ (only when the length fluctuations are present).  

For the symmetric case $A=B$ Eq.~(\ref{eq:gen2}) simplifies into
\begin{widetext}
\begin{eqnarray}
C(\eta_1, \eta_2) &=& 1 +  \frac{1}{\br{N_+}^2} \label{eq:gensym2} \\ &\times&
                            \Big\{ \br{N_+} \left [ r( 1 - {u_1 u_2} -  |u_1-u_2|) +s(\omega)(1+r + (1-r)u_1 u_2  - r |u_1-u_2| ) \right ] + {\rm var}(N_+)  + {\rm var}(N_-) u_1 u_2 
                           \Big\}, \nonumber
\end{eqnarray}
where we note the additional symmetry $(\eta_1 \to -\eta_1$, $\eta_2 \to -\eta_2)$.
For the strongly asymmetric case ($N_A \gg N_B$) we get
\begin{eqnarray}
&& C(\eta_1, \eta_2) = 1 + \frac{{\rm var}(N_A)}{\br{N_A}^2}+ \frac{s(\omega)}{\br{N_A}}+ r  \frac{[1+s(\omega)][1-u_1u_2-|u_1-u_2|]}{\br{N_A} (1 + u_1)(1+u_2)} +\mathcal{O}(N_B/N_A), \label{eq:genasym2}
\end{eqnarray}
\end{widetext}

In the limiting case of no length fluctuations ($r=0$) and no overlaid distribution ($s(\omega)=0$) for symmetric collisions, we recover from Eq.~(\ref{eq:gensym2})
the result by Bzdak and Teaney~\cite{Bzdak:2012tp},
\begin{eqnarray}
C(\eta_1,\eta_2) = 1 + \frac{{\rm var}(N_+) + {\rm var}(N_-) u_1 u_2 }{\br{N_+}^2}, \label{eq:btr}
\end{eqnarray}
which demonstrates that the fluctuation of $N_A$ vs $N_B$ induces the  $\eta_1 \eta_2$ structure in $C(\eta_1, \eta_2)$. 

We remark that in the Glauber model one has approximately  ${\rm var}(N_+)\sim \br{N_+}$ and  ${\rm var}(N_-)\sim \br{N_+}$, hence
Eq.~(\ref{eq:gensym2}) leads to the approximate $1/\br{N_+}$  scaling of $C(\eta_1,\eta_2)-1$ for the symmetric case. 
Similarly, from Eq.~(\ref{eq:genasym2}) we get the scaling $1/\br{N_A}$ for the strongly asymmetric case.

\section{Longitudinal correlations in the wounded nucleon model \label{sec:wounded}}

To see the importance of the length fluctuations, we have carried out a {\tt GLISSANDO}~\cite{Broniowski:2007nz,*Rybczynski:2013yba} simulation in the wounded nucleon model 
for Pb+Pb collisions at 2.76~TeV. In this case $N_A$ and $N_B$ are the wounded nucleons in the $A$ and $B$ nuclei.
The simulation is needed solely to obtain the statistical averages  $\br{N_+}$, ${\rm var}(N_+)$, and ${\rm var}(N_-)$, as the dependence on 
$\eta_1$ and $\eta_2$ is analytic. 

Some explanation is needed concerning the overlaid distribution of $\omega$. The quantity  $\int_{-Y}^Y d\eta_1 \int_{-Y}^Y d\eta_2 C(\eta_1,\eta_2)$
measures the overall multiplicity fluctuation in the experimental pseudorapidity coverage. Therefore, when comparing the models with and without 
length fluctuations we should make this quantity equal. This requirement yields, with the help of Eq.~(\ref{eq:gensym2}), the condition
\begin{eqnarray}
 s(\omega)_{\rm length~fl.}  = \frac{ 3 s(\omega)_{\rm no~ length~fl.}+2 \frac{Y}{y_p}-3}{6-2  \frac{Y}{y_p}}.  \label{eq:ome}
\end{eqnarray}
This equation shows, according to the expectations, that without the length fluctuations we must add more fluctuations in $\omega$ to obtain the same 
amount of the overall multiplicity fluctuations. In Refs.~\cite{Bozek:2013uha,Bozek:2015bna} we have checked that the overlaid 
$\Gamma$ distribution describes properly the multiplicity distribution in p-Pb collisions at 5.02~TeV.   
The parameters for the model without the length fluctuations yield $s(\omega)\sim 1$, hence we take $s(\omega)_{\rm no~ length~fl.}=1$, and, correspondingly to Eq.~(\ref{eq:ome}), 
$s(\omega)_{\rm length~fl.}=Y/(3y_p-Y) \simeq 0.1$. Such a low value for the model with the length fluctuations present is consistent 
with the observation made in Ref.~\cite{Monnai:2015sca}, where no overlaid distribution was needed to reproduce the multiplicity spectra.

\begin{figure}
\begin{center}
{\includegraphics[width=0.4\textwidth]{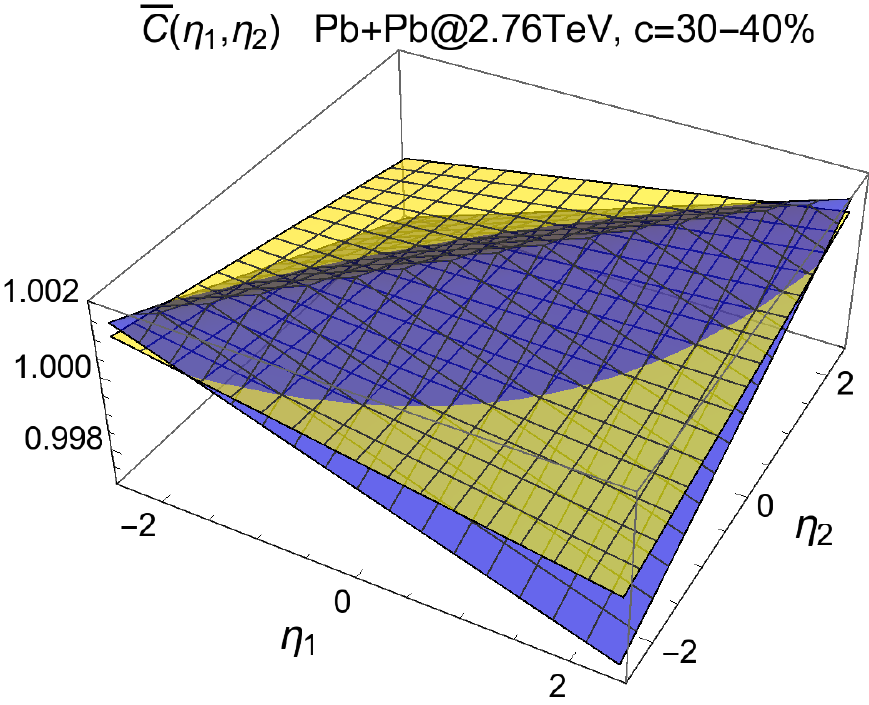}}
\end{center}
\vspace{-5mm}
\caption{(Color online) Correlation function $\overline{C}(\eta_1,\eta_2)$ for Pb-Pb collisions at 2.76~TeV for centrality $c=30-40\%$, obtained from 
Eqs.~(\ref{eq:Cbar},\ref{eq:gensym2}). The flat (light color) sheet corresponds to the case without the length fluctuations, whereas the sheet with the elongated maximum (darker color) 
corresponds to the model with the length fluctuations. \label{fig:c3040}}
\end{figure}

The result for the normalized correlation function (\ref{eq:Cbar}) for Pb-Pb collisions at 2.76~TeV for a sample centrality $c=30-40\%$ 
is shown in Fig.~\ref{fig:c3040}. We note a vivid difference between the models without and with the length fluctuations. Whereas the former case 
shows a rather flat structure, including the $\eta_1 \eta_2$ term, the latter displays an elongated maximum, due to the $|\eta_1-\eta_2|$ structure 
resulting from the length fluctuations. 
 
\begin{figure}
\begin{center}
{\includegraphics[width=0.4\textwidth]{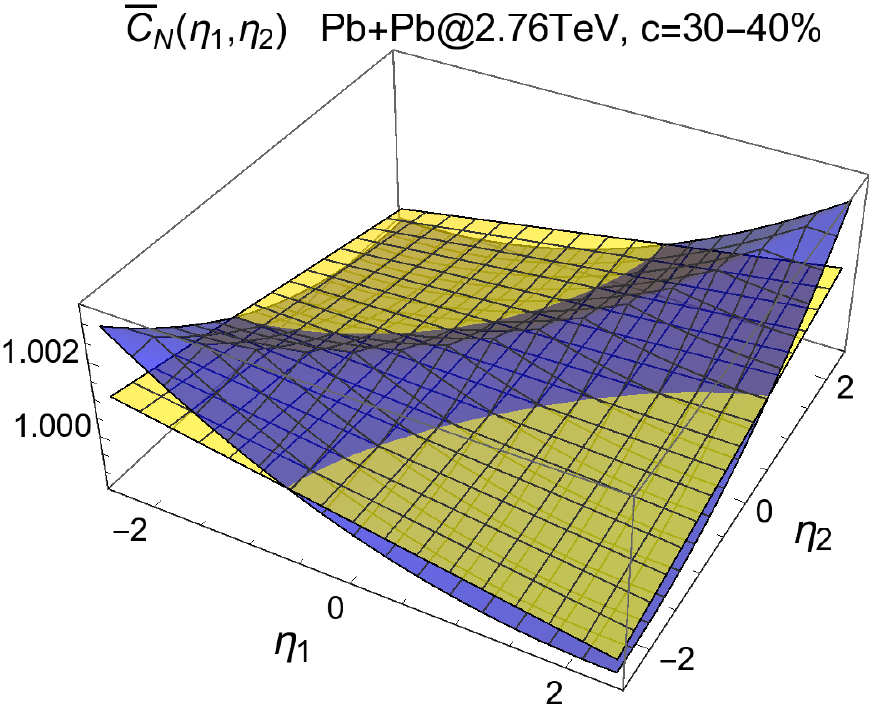}}
\end{center}
\vspace{-5mm}
\caption{(Color online) Same as Fig.~\ref{fig:c3040} but for the function  $\overline{C}_N(\eta_1,\eta_2)$. \label{fig:c3040N}}
\end{figure}

In Fig.~\ref{fig:c3040N} we show the corresponding function  $\overline{C}_N(\eta_1,\eta_2)$.  
We note the generation of the ridge with a saddle. The shape is simply caused by the definition (\ref{eq:CN}); as the marginal distributions $C_p(\eta)$ have a maximum at $\eta=0$, the 
denominator in Eq.~(\ref{eq:CN}) relatively enhances the part of the plot at larger values of $\eta_1$ and $\eta_2$. 
Hence the generation of the saddle in the ridge (as seen in the experiment~\cite{ATLAS:2015kla}) is natural from our expressions. 

We note that the shape of the correlation functions with the length fluctuations displayed in Figs.~\ref{fig:c3040} and \ref{fig:c3040N} 
is qualitatively very similar to the experimental results, cf. Fig.~1 and 3 in Ref.~\cite{ATLAS:2015kla}. Also, the  values are comparable. At other 
centralities the shape of our correlation functions is as in Figs.~\ref{fig:c3040} and \ref{fig:c3040N}, but the size reflects the approximate $1/\br{N_+}$ scaling.

\begin{figure}
\begin{center}
{\includegraphics[width=0.4\textwidth]{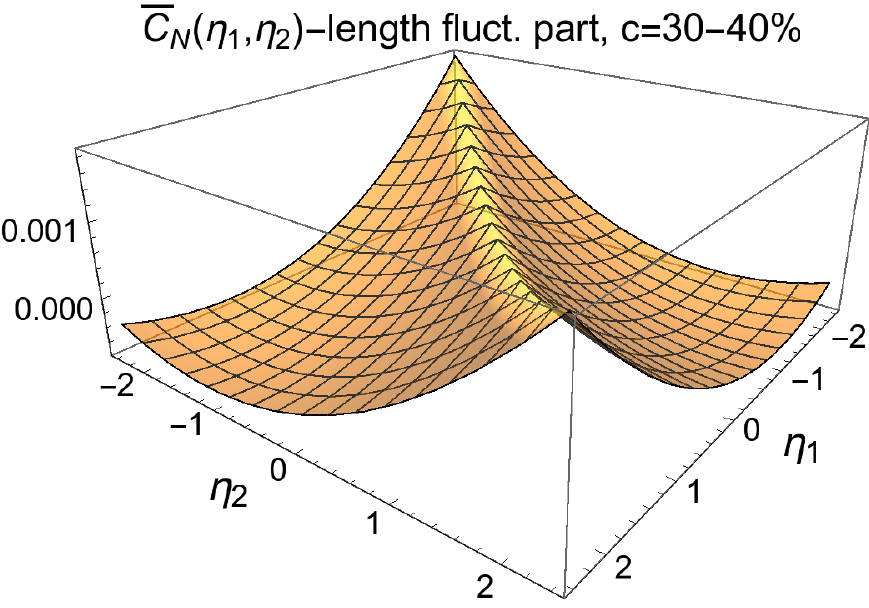}}
\end{center}
\vspace{-5mm}
\caption{(Color online) The difference of the correlation function from Fig.~(\ref{fig:c3040N}) with and without length fluctuations.  \label{fig:grz}}
\end{figure}

\begin{figure}
\begin{center}
{\includegraphics[width=0.4\textwidth]{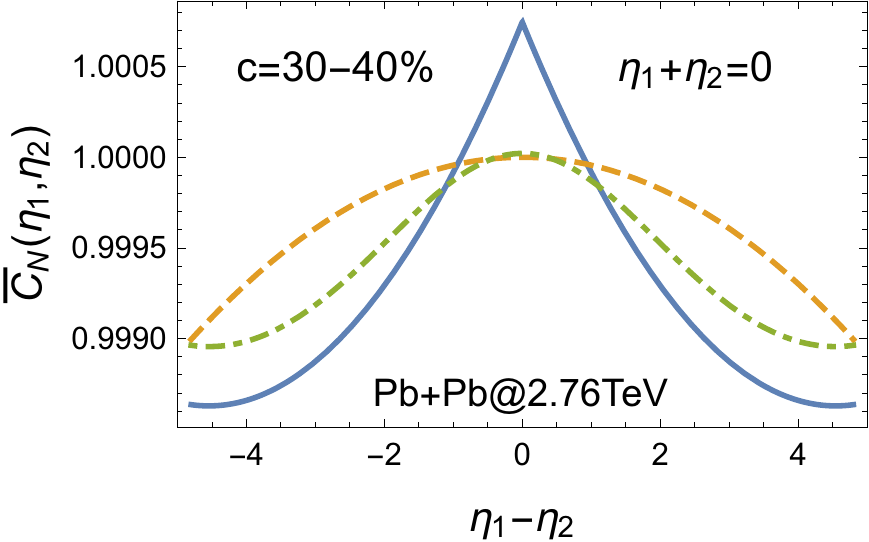}}
\end{center}
\vspace{-5mm}
\caption{(Color online) Sections of $\overline{C}_N(\eta_1,\eta_2)$ along the line $\eta_1+\eta_2=0$. The solid (dashed) lines 
correspond to the model with (without) the length fluctuations plotted in Fig.~\ref{fig:c3040N}. The dot-dashed curve shows 
the result of smearing of the model with the length fluctuations with $\sigma_\eta=1$, as described in Sec.~\ref{sec:tohadrons}. \label{fig:cut}}
\end{figure}

To focus on the contribution from the length fluctuations, in Fig.~\ref{fig:grz} we display the difference of the cases including and excluding this effect.
We note that the half-width of the ridge structure in the $\eta_1-\eta_2$ coordinate is about 2 units of pseudorapidity. The effect is clearly seen in 
Fig.~\ref{fig:cut}, where we show the sections of Fig.~(\ref{fig:c3040N}) along the line $\eta_1+\eta_2=0$.

\section{The $a_{nm}$ coefficients}

Simple analytic expressions for the $a_{nm}$ coefficients for symmetric collisions may be obtained from Eq.~(\ref{eq:gensym2}). 
We find (for $n,m>0$)
\begin{eqnarray}
a_{nn}&=& \frac{\frac{{\rm var}(N_-)}{\br{N_+}}+(1-r)s(\omega)-r}{6 \br{N_+}}\frac{Y^2}{y_b^2} \delta_{n1} \nonumber \\
&+& r \frac{s(\omega)+1}{(2n-1)(2n+3) \br{N_+}}\frac{Y}{y_b}, \label{eq:ann} \\  \nonumber \\
a_{n,n+2} &=&a_{n+2,n} = \nonumber \\ &=& -r \frac{s(\omega)+1}{2(2n+3)\sqrt{(2n+1)(2n+5)} \br{N_+}}\frac{Y}{y_b}, \nonumber
\end{eqnarray}
with all remaining combinations of $n$ and $m$ yielding $a_{nm}=0$. The first two terms in $a_{nn}$ originate 
from the $\eta_1 \eta_2$ piece, while the other terms come from $|\eta_1-\eta_2|$. We note several facts: 
\begin{enumerate}
 \item In the model without length fluctuations ($r=0$) 
we only have $a_{11} \neq 0$, which complies to Eq.~(\ref{eq:btr}). 
 \item The coefficients scale as $1/\br{N_+}$. For $a_{11}$ there may be slight departures from this scaling from the term 
 ${\rm var}(N_-)/\br{N_+}$, whereas for other coefficients the scaling is exact.
 \item The coefficients drop with the value of the rank $n$, with the behavior $1/n^2$ at large $n$.
\end{enumerate}

\begin{figure}
\begin{center}
{\includegraphics[width=0.4\textwidth]{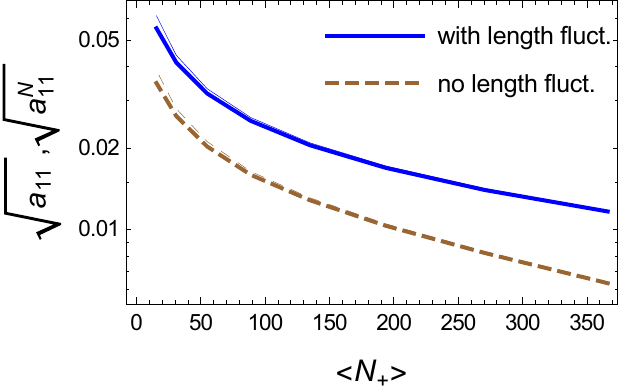}}
\end{center}
\vspace{-5mm}
\caption{(Color online)  $\sqrt{a_{11}}$ and $\sqrt{a_{11}^N}$ (thin and thick lines, respectively) in the wounded nucleon model for Pb+Pb collisions at 2.76~TeV. 
\label{fig:a11}}
\end{figure}

Analogous analytic expressions for the $a_{nm}^N$ coefficients are lengthy, so we present them numerically. Their values are close to the 
$a_{nm}$ coefficients, with larger departure at low values of $\br{N_+}$. 
This feature is seen for $a_{11}$ and $a_{11}^N$ from Fig.~\ref{fig:a11}. We note that the difference between the model with and without 
length fluctuations is large (about a factor of 2).

From the occurrence of $\eta^2$ in Eq.~(\ref{eq:cpsym}) it is clear that in the symmetric case the coefficients $a_{nm}^N$ are no longer limited to the tridiagonal structure
of Eq.~(\ref{eq:ann}), but all combinations with $n$ and $m$ different by multiples of 2 are possible. However, the coefficients with $|n-m|>2$ 
are strongly suppressed with powers $1/\br{N_+}^{|m-n|/2}$. 

For the asymmetric case all combinations of $n$ and $m$ are possible, with a similar effect of suppression as one departs from the diagonal. 
For $a_{11}$ we find
\begin{eqnarray}
 a_{11}&=&r (s(\omega)+1) y_p \times \label{eq:a11asym} \\
&& \frac{3 y_p \left({Y}^2-{y_p}^2\right) {\rm arctanh}
   \left(\frac{{Y}}{{y_p}}\right)+3 {y_p}^2 {Y}-2 {Y}^3}{2 \br{N_A}{Y}^4}  \nonumber \\
&=&r \frac{(s(\omega)+1) {Y}}{5 \br{N_A} {y_p}}+r\frac{3 (s(\omega)+1) {Y}^3}{35 \br{N_A}
   {y_p}^3}+\mathcal{O}(Y^4/y_p^4).    \nonumber
\end{eqnarray}

\subsection{Universal scaling}

We notice that to leading order in $Y/y_p$, i.e., for small acceptance compared to the beam rapidity, we have for both the
symmetric and asymmetric cases the universal behavior
\begin{eqnarray}
a_{11}=  r\frac{(s(\omega)+1)}{5} \frac{Y}{y_p} \frac{1}{\br{N}} + \dots, \label{eq:scale}
\end{eqnarray}
where the dots indicate higher-order terms in $Y/y_p$. Here $N$ denotes $N_+$ for the symmetric case, and $N_A$ for the asymmetric case, i.e., the 
total number of sources. Since for the ATLAS coverage $Y/y_p \sim 30\%$, higher-order terms can be neglected and the universal (i.e., reaction-independent) 
$1/\br{N}$ scaling of Eq.~(\ref{eq:scale}), with the same prefactor, holds to a
 good accuracy.

In the scaling equation (\ref{eq:scale}) the quantity $\br{N}$ represents the average number of independent longitudinally extended sources (strings, flux tubes),
emitting particles in the experimental acceptance window. In the illustrative calculation of the preceding section we have used wounded nucleons as sources. 
However, it should be noted that the same expressions apply if wounded quarks are connected to the decaying flux tubes. 
The ATLAS experiment finds the $1/N_{\rm ch}$  scaling for the $a_{nm}$ coefficients~\cite{ATLAS:anm}, where $N_{\rm ch}$ denotes the number of observed charged hadrons,
with approximately the same prefactor for Pb-Pb, p-Pb, and p-p collisions when the short-range correlations are removed (cf. Fig.~14 of Ref.~\cite{ATLAS:anm}).
If the scaling between the number of sources and the multiplicity of charged particles is linear, 
\begin{eqnarray}
N \sim N_{\rm ch}, \label{eq:nnch}
\end{eqnarray}
then the scaling $a_{nm}\propto 1/\br{N}$ is transformed into the scaling $a_{nm} \propto 1/\br{N_{ch}}$, exactly as observed in the experimental data.

We note that the proportionality (\ref{eq:nnch}), with $N$ as the number of wounded quarks, was checked~\cite{Eremin:2003qn,Adler:2013aqf} to hold well 
at the RHIC energies. On the other hand, this proportionality does not hold when $N$ stands for the number of wounded nucleons.
Thus we conclude that Eq.~(\ref{eq:scale}) together with the experimental $1/N_{\rm ch}$  scaling for the $a_{nm}$ coefficients 
conforms to the wounded quark picture of the high-energy nuclear reactions. 

\subsection{Model vs data}

We now pass to comparing our model results to the data. Since the later ATLAS analysis~\cite{ATLAS:anm} removes the short-distance component from the correlation 
function with a rather involved procedure, difficult to repeat in a model calculation, we resort to the preliminary data from Ref.~\cite{ATLAS:2015kla} for the pair of 
particles of the same charge. These correlation functions are not strongly sensitive to resonance decays.

\begin{figure}
\begin{center}
{\includegraphics[width=0.4\textwidth]{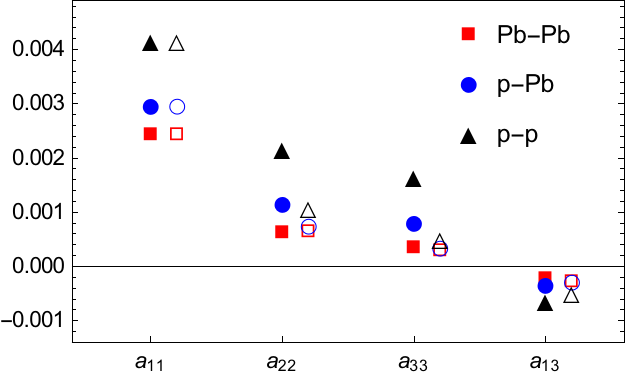}}
\end{center}
\vspace{-5mm}
\caption{(Color online) The ${a_{nm}}$ coefficients evaluated in the model (open symbols) and compared to the preliminary ATLAS data (filled symbols) 
for the same-charge hadron pairs, takes from Fig.~7 of Ref.~\cite{ATLAS:2015kla} for 
$N_{\rm ch}=100$-$120$. The experimental values of $a_{11}$ were used to determine the proportionality between  $\br{N_{\rm ch}}$ and the average number of sources $\br{N}$
for each reaction.
\label{fig:nm}}
\end{figure}

We can use the experimental values for the $a_{11}$ coefficients to fix the proportionality constant in Eq.~(\ref{eq:nnch}). 
To reproduce the data for $a_{11}$ in  Fig.~\ref{fig:nm} we have assumed that $\br{N_{\rm ch}}=4.7 \br{N_+}$ for Pb-Pb collisions at 2.76~TeV,   $\br{N_{\rm ch}}=5.1 \br{N_A}$ for p-Pb collisions at 5.02~TeV, and 
$\br{N_{\rm ch}}=8.1 \br{N_+}$ for p-p collisions at 13~TeV. 
These coefficients have the interpretation of the average number of observed charged hadrons per source, and have reasonable values. 
They roughly reflect the growth of the  multiplicity in p-p collisions with energy. Moreover,
it should be noted that the same multiplicity classes in different collision systems correspond 
to a different percentile of the inelastic cross section. In high multiplicity 
p-p events one triggers on events with a higher multiplicity of hadrons per source. A detailed
modeling of the  multiplicity classes would require a Monte Carlo modeling and additional assumptions concerning the distribution of sources and the number 
of charged particles per source. These issues are outside the scope of the analytic calculation presented in this paper. 

We note from Fig.~\ref{fig:nm} that
the higher-rank coefficients are predicted to be significantly smaller than $a_{11}$. Whereas the case of Pb-Pb is 
well reproduced for higher $n$ and $m$, the splitting between Pb-Pb and p-Pb or p-p is too small.
The higher-order coefficients are expected to be more sensitive to the details of production mechanism that are not taken into account in 
our analytic calculation, such as 
a non-uniform distribution of flux-tube ends, matter evolution in rapidity, difference between rapidity and pseudorapidity, or hadronization effects. 
A strong sensitivity of higher order $a_{nm}$ coefficients to the details of the dynamics has been noticed in Ref.~\cite{Monnai:2015sca}.

\section{From initial state to final hadrons \label{sec:tohadrons}} 

Our analytic model uses several simplifying assumptions.
We comment shortly about these issues and make an estimate 
of the most important effect.

The dynamics of particle creation acts differently in elementary collisions and in nuclear collisions. Hadronization 
in a small system is severely constrained by the local energy and charge conservation requirement~\cite{Adam:2015gda}, which 
leads to modifications of the correlation functions.
Charge conservation generates a peak in the two-particle correlation as small pseudorapidity separations \cite{Jeon:2001ue,*Bozek:2012en}.
 The energy conservation in the string 
fragmentation mechanism gives a reduction of the emission probability for particles with similar rapidities.
If a large collectively expanding source is created, the energy conservation effects
are reduced. On the other hand, the longitudinal expansion of the fireball 
may induce a rescaling of the rapidity variable between the initial state 
and the matter at freeze-out. This would lead to some reduction of the correlation 
coefficients $a_{nm}$ in the final state as compared to the initial state.

The transformation from the density $\rho(\eta_1,\eta_2)$ into observed 
particles involves a hadronization mechanism. In elementary processes,
it could be modeled as string decay or quark coalescence. In nuclear 
collisions, individual hadrons appear at freeze-out through  
thermal emission from fluid elements. In the first case, particle emission
requires color and momentum exchanges with the rest of the system 
in order to fulfill the local conservation laws. In the case of
the emission from a hot fireball, hadron rapidities are washed out as compared to the fluid rapidity due to the thermal component in the momenta.
A similar washing out of the distribution is expected from the decays of resonances.

\begin{figure}
\begin{center}
{\includegraphics[width=0.4\textwidth]{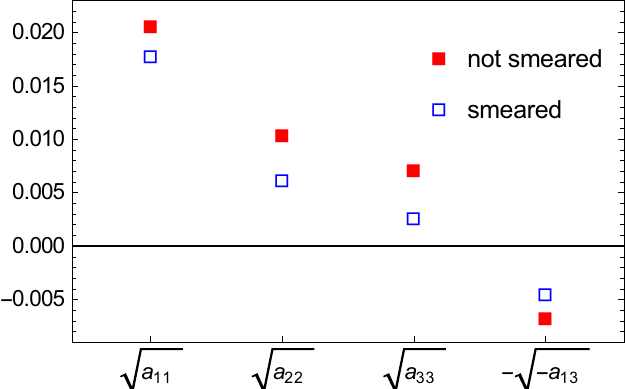}}
\end{center}
\vspace{-5mm}
\caption{(Color online) Square root of the ${a_{nm}}$ coefficients evaluated in the model without smearing  (filled symbols)  and with a Gaussian smearing of with $\sigma_\eta=1$ (empty symbols). 
Pb-Pb collisions at 2.76~TeV, wounded nucleon model with {\tt GLISSANDO}, centrality \mbox{$c=30-40\%$}.
\label{fig:nmsmear}}
\end{figure}

Irrespective of the details of the mechanism of particle production, the 
pseudorapidity distribution of the observed particles is different than the two-particle density discussed so far.
The effects can be approximately estimated as a convolution of the numerator of Eq.~(\ref{eq:C})
with Gaussian form factors,
\begin{equation}
S_{\rm sm}(\eta_1,\eta_2)=\int d\eta_1'\int d\eta_2' g(\eta_1,\eta_1')
g(\eta_2,\eta_2') S(\eta_1',\eta_2'),  \label{eq:smear}
\end{equation}
and similarly for the denominator $B(\eta_1,\eta_2)$. 
We take the following form of the smearing function 
\begin{equation}
g(\eta,\eta^{'})=\frac{1}{\sqrt{2\pi} \sigma_\eta} e^{-\frac{(\eta-\eta^{'})^2}{2\sigma_\eta^2}},
\label{eq:smearf}
\end{equation}
with $\sigma_\eta=1$,
which corresponds to a rather large smearing width of $\sqrt{2}$ for the difference $\eta_1-\eta_2$.
We notice from the results shown in Fig.~\ref{fig:nmsmear} that the smearing reduces the values of the $a_{nm}$ coefficients. As expected,  
the effect is stronger for higher-rank coefficients. 

\begin{figure}
\begin{center}
{\includegraphics[width=0.45\textwidth]{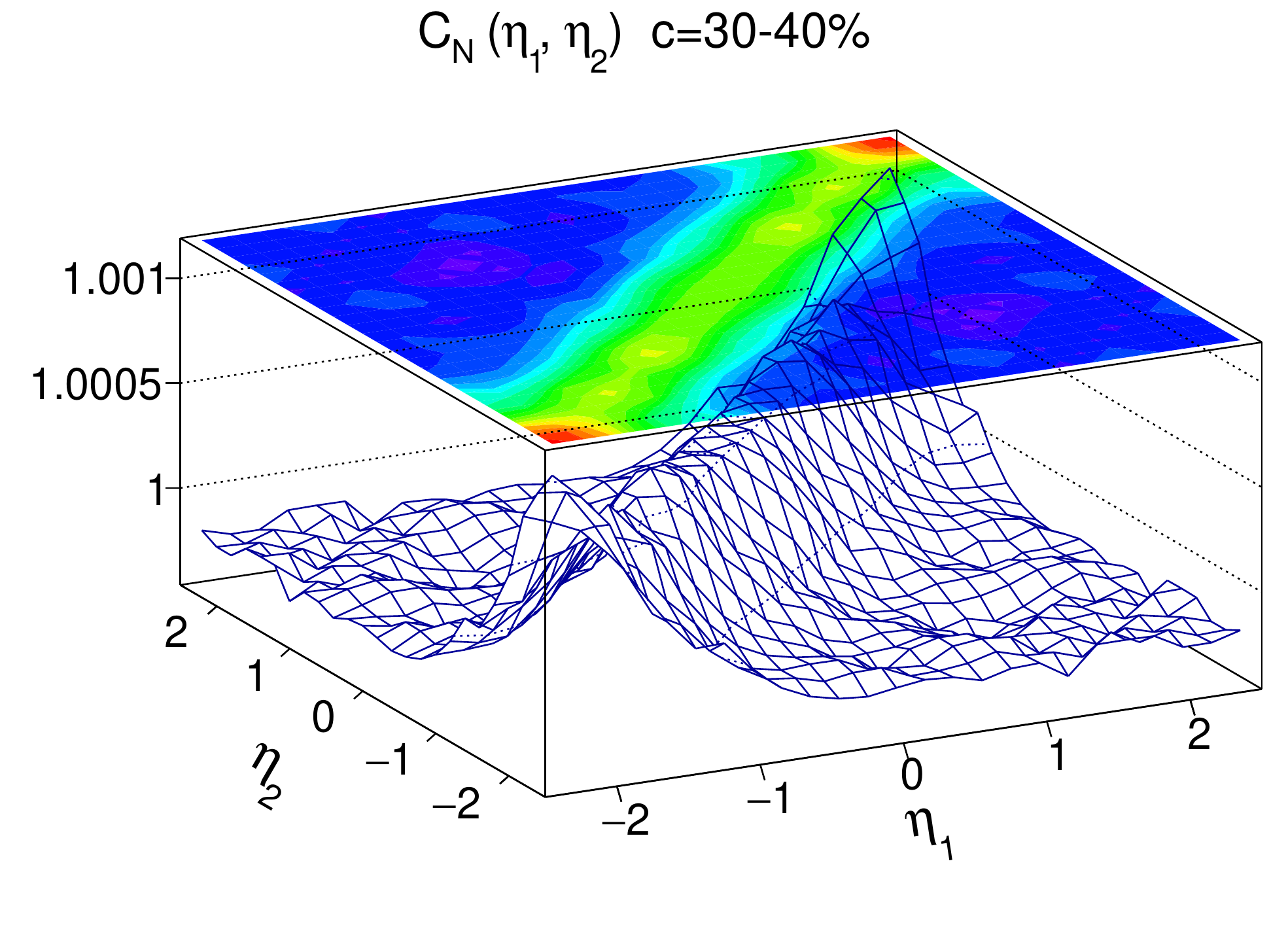}}
\end{center}
\vspace{-7mm}
\caption{(Color online)
Contribution to the correlation function $C_N(\eta_1,\eta_2)$ from the resonance decays, reprinted from Fig.~2 of Ref.~\cite{Bozek:2015tca}.
\label{fig:reso}}
\end{figure}

Finally, resonance decays are a significant source of two-particle correlations among produced hadrons of opposite charge.
In Fig.~\ref{fig:reso} we recall the result of our analysis of Ref.~\cite{Bozek:2015tca} carried out in the 
wounded nucleon model without fluctuating ends.
Comparing Fig.~\ref{fig:reso} to Fig.~\ref{fig:c3040N}, we note a significant relative size of the component from resonance decays.

\section{Conclusions \label{sec:concl}}

The mechanism of energy deposition and particle creation in high-energy
collisions in the longitudinal direction has been a subject of many 
recent studies. In this paper, we have investigated the two-particle correlations
in rapidity as measured by the ATLAS Collaboration~\cite{ATLAS:2015kla,ATLAS:anm}. 
A nice aspect of the proposed correlation measures is that they may be obtained analytically in a simple model 
based on independent longitudinally extended sources.
Our model  exemplifies the most important characteristics of a broad 
class of realistic models based on the decays of strings or flux tubes, while remaining simple 
enough to allow for the derivation of analytic formulas for the correlation
function and for the coefficients of its expansion in orthogonal polynomials.

Our model of the initial energy (or entropy) deposition from flux tubes 
is a mechanism describing the formation of the initial
state in the collisions, with the distribution involving two-particle correlations. 
The model has been shown previously to describe fairly well the decorrelation of the harmonic flow event-planes at different pseudorapidities, also 
in p-Pb collisions~\cite{Bozek:2015bna}.
The correlations in the energy deposition from the early stage of the collision can be 
transmitted into final hadron distribution via different scenarios. In elementary 
collisions, the flux tubes decay directly into hadrons at hadronization, 
with subsequent resonance decays. In nuclear collisions (or, perhaps, highest-multiplicity p-p collisions), a fireball is formed that expands collectively 
and hadrons are emitted at freeze-out. To a good approximation, our analysis applies to both scenarios, as the hydrodynamic evolution
of the fireball in the longitudinal direction for central rapidities is moderate. 

The flux tubes are formed independently from one another from excited color charges in the  target and the projectile.
Our model explicitly reveals the two basic mechanism of generating fluctuations in rapidity: Firstly, 
an event-by-event asymmetry of the number of excited charges in the target and projectile 
leads to event-by-event fluctuations of the density and gives a correlation function of the form $C(\eta_1,\eta_2)= 1+ A \eta_1\eta_2 + \dots$, 
as already noticed by Bzdak and Teaney~\cite{Bzdak:2012tp}. The second mechanism originates from the random
positions in rapidity of the end-points of the decaying sources (length fluctuations). We find that this effect generates the dominant contribution to the 
correlation function, roughly two times larger than the Bzdak-Teaney term. Both terms scale approximately as $1/\br{N}$, i.e., are inversely proportional 
to the number of sources. With the feature that the number of sources is proportional to the final multiplicity of charged hadrons, 
the model explains the experimental observation of a similar value of the $a_{11}$ coefficients in different 
systems at the same multiplicity~\cite{ATLAS:anm}, as well as its $1/\br{N_{\rm ch}}$ scaling,

Our analytic expressions for the correlation function and the $a_{nm}$ coefficients display the dependence on statistical measures of the sources and the 
overlaid distribution, on the assumed model of the end-point fluctuation, on the width of the experimental acceptance window in pseudorapidity, 
and on the indices $n$ and $m$.  One should bear in mind that the derived formulas are modified by a number of omitted and potentially 
relevant effects, in particular hadronization and resonance decays. 
We have crudely estimated these effect by smearing the correlation function with Gaussian form factors, which leads to a 
sizable reduction of the $a_{nm}$ coefficients, stronger for 
larger $n$ and $m$. Our smearing procedure may be thought of as a way to remove some of the short-distance correlations.
The resonance decays, analyzed via numerical simulations in  Ref.~\cite{Bozek:2015tca}, form a significant explicit contribution to the correlation function.
It may be significantly reduced when the correlations of pairs of the same charge are considered.

\begin{acknowledgments}

Research supported by the Polish Ministry of Science and Higher Education, (MNiSW), by the Polish National
Science Centre grants DEC-2012/05/B/ST2/02528.

\end{acknowledgments}

\appendix

\section{Derivation of $C(\eta_1,\eta_2)$ in the longitudinally-extended source model \label{sec:lesm}}

The formula for the event-by-event averaged distribution of entropy is 
\begin{eqnarray}
&& \hspace{-7mm} \br{\rho(\eta_1.\eta_2)} = \nonumber \\
&=& \br{N_A}  \br{f_A(\eta_1,\eta_2)} + \br{N_A(N_A-1)} \br{f_A(\eta_1)} \br{f_A(\eta_2)} \nonumber \\
                                      &+& \br{N_B}  \br{f_B(\eta_1,\eta_2)} + \br{N_B(N_B-1)} \br{f_B(\eta_1)} \br{f_B(\eta_2)} \nonumber \\
                                      &+& \br{N_A N_B} \left [  \br{f_A(\eta_1)} \br{f_B(\eta_2)}+\br{f_B(\eta_1)} \br{f_A(\eta_2)} \right ], \nonumber \\
\label{eq:rhoformula}
\end{eqnarray}
where the first term corresponds to the situation where both bins at $\eta_1$ and $\eta_2$ are fed from the production from the 
same source associated to nucleus $A$, the second term corresponds to production from two different sources belonging to $A$, 
the third and fourth term describe the analogous emission from the sources associated to $B$, finally, the last term corresponds to the emission 
from one source belonging to $A$ and the other to $B$. 
The functions $\br{f_{A,B}(\eta_1,\eta_2)}$ and $\br{f_{A,B}(\eta_{1,2})}$ are the event-averaged two-body and one-body emission profiles, 
constructed below.

With a simple rearrangement one may rewrite the above equation in the form
\begin{eqnarray}
&& \hspace{-7mm} \br{\rho(\eta_1,\eta_2)} = \nonumber \\
&=& \br{N_A}  {{\rm cov}_A(\eta_1,\eta_2)} + \br{N_A^2} \br{f_A(\eta_1)} \br{f_A(\eta_2)} \nonumber \\
                                      &+& \br{N_B}  {{\rm cov}_B(\eta_1,\eta_2)} + \br{N_B^2} \br{f_B(\eta_1)} \br{f_B(\eta_2)} \nonumber \\
                                      &+& \br{N_A N_B} \left [ \br{f_A(\eta_1)} \br{f_B(\eta_2)}+\br{f_B(\eta_1)} \br{f_A(\eta_2)} \right ] , \nonumber \\
\end{eqnarray}
where 
\begin{eqnarray}
{\rm cov}_i(\eta_1,\eta_2)= \br{f_i(\eta_1,\eta_2)} -  \br{f_i(\eta_1)} \br{f_i(\eta_2)}, \;\; i=A,B. \nonumber \\
\end{eqnarray}

For the event-by-event averaged one-body densities we simply have
\begin{eqnarray}
\br{\rho(\eta)} = \br{N_A}  \br{f_A(\eta)} + \br{N_B}  \br{f_B(\eta)}.
\end{eqnarray}

In the reference frame where the $A$ and $B$ nuclei move with equal and opposite velocities, we may decompose the emission 
profiles into parts symmetric and antisymmetric in spatial pseudorapidity. 
\begin{eqnarray}
&& \br{f_A(\eta)}=\br{f_s(\eta)}+\br{f_a(\eta)}, \;  \br{f_B(\eta)}=\br{f_s(\eta)}-\br{f_a(\eta)}, \nonumber \\
&& \br{f_s(\eta)}=\br{f_s(-\eta)}, \;\;\; \br{f_a(\eta)}=-\br{f_a(-\eta)}. \label{eq:symasym}
\end{eqnarray}
Then, after elementary transformations,
\begin{widetext}
\begin{eqnarray}
C(\eta_1, \eta_2) = 1 &+&  \frac{1}{\left[ \br{N_+}\br{f_s(\eta_1)}\!+\! \br{N_-}\br{f_a(\eta_1)} \right] \left[ \br{N_+}\br{f_s(\eta_2)}\!+\! \br{N_-}\br{f_a(\eta_2)} \right]} \times  \nonumber \\
                            &\times& \Big\{ \br{N_A} {\rm cov}_A(\eta_1,\eta_2) + \br{N_B} {\rm cov}_B(\eta_1,\eta_2) 
                            + {\rm var}(N_+) \br{f_s(\eta_1)}\br{f_s(\eta_2)} + {\rm var}(N_-) \br{f_a(\eta_1)}\br{f_a(\eta_2)} \nonumber \\
                            &+& {\rm cov}(N_+,N_-) \left [\br{f_s(\eta_1)} \br{f_a(\eta_2)}+ \br{f_a(\eta_1)} \br{f_s(\eta_2)} \right ] \Big\}, \label{eq:gen}
\end{eqnarray}
where ${\rm cov}(N_+,N_-) = {\rm var}(N_A)-{\rm var}(N_B)$.
For the special case of symmetric collisions, where $\br{N_A}=\br{N_B}$ and ${\rm var}(N_A)={\rm var}(N_B) $, Eq.~(\ref{eq:gen}) simplifies into
\begin{eqnarray}
C(\eta_1, \eta_2) &=& 1 + \frac{1}{\br{N_+}^2 f_s(\eta_1) f_s(\eta_2)} \times \nonumber \\
                            &\times& \Big\{ \br{N_A} {\rm cov}_A(\eta_1,\eta_2) + \br{N_B} {\rm cov}_B(\eta_1,\eta_2) + 
                            {\rm var}(N_+) \br{f_s(\eta_1)} \br{f_s(\eta_2)} + {\rm var}(N_-) \br{f_a(\eta_1)} \br{f_a(\eta_2)} \Big\}.  \label{eq:gensym}
\end{eqnarray}

Another limiting case is for $N_A \gg N_B$, when
\begin{eqnarray}
C(\eta_1, \eta_2) &=& 1 + \frac{1}{\br{N_A}^2  \br{f_A(\eta_1)}\br{f_A(\eta_2)}}  
 \Big\{ \br{N_A} {\rm cov}_A(\eta_1,\eta_2) + {\rm var}(N_A)  \br{f_A(\eta_1)}\br{f_A(\eta_2)} \Big\}  + \mathcal{O}(N_B/N_A). \label{eq:genasym}
\end{eqnarray}
\end{widetext}

A general remark may be made here. We note from Eq.~(\ref{eq:gen}-\ref{eq:genasym}) that the correlations originate from two kinds of effects: the correlations 
from the emission off a single source (the terms with ${\rm cov}_{A,B}(\eta_1,\eta_2) $) and the fluctuations of the numbers of sources (the terms with variances). 
Therefore, as noticed already in Ref.~\cite{Bzdak:2012tp}, even if ${\rm cov}_{A,B}(\eta_1,\eta_2)=0$, we have a  nontrivial correlation function $C(\eta_1,\eta_2)$.

\section{Fluctuating ends \label{sec:ends}}

Bia\l{}as-Czy\.z \cite{Bialas:2004su} have shown that the d+Au collisions at RHIC may be understood within the wounded nucleon model if the 
average emission profiles have the form 
\begin{eqnarray} 
\br{f_A(\eta)}=h(\eta) t(\eta;y_b),  \;\; \br{f_B(\eta)}=h(\eta)t(-\eta;y_b), \nonumber \\ \label{eq:tri}
\end{eqnarray}
where 
$y_b$ is a parameter of the order of the rapidity of the beam,
\begin{eqnarray}
t(\eta;y_b)=\left \{ \begin{array}{cl} 0 & {\rm for~} \eta<-y_b\\ \frac{y_b+\eta}{2 y_b} &  {\rm for~} -y_b \le \eta \le y_b \\ 1  &  {\rm for~} y_b < \eta \end{array} \right. ,  \label{eq:tria}
\end{eqnarray}
and $h(\eta)$ is a suitable chosen function, symmetric in $\eta$, which turns out to be flat near the origin~\cite{Bozek:2011ua},
\begin{eqnarray}
h(\eta)\simeq{\rm const}\;\;\;{\rm for~} -2.4<\eta<2.4 \label{eq:flat}
\end{eqnarray}
at the LHC collision energies (as we work up to a multiplicative constant, we may take $h(\eta)=1$).
Note the presence of the 
antisymmetric part in $ t(\eta;y_b)$  which describes the fact that the source from $A$ deposits entropy mostly forward, and the source from $B$ mostly backward in $\eta$.
Parametrization~(\ref{eq:tri}) has been later used in numerous papers \cite{Adil:2005bb,*Bozek:2010bi} as a working model for the initial distributions.

We now step up in the model building and investigate models where the sources have randomly distributed ends. The constraint is that they must reproduce the phenomenologically successful Eq.~(\ref{eq:tri}).

\subsection{Single-end fluctuations \label{sec:single}}

We consider first a simple model where the source has one end attached to the nucleus $A$ or $B$ (it is placed at pseudorapidity $y_b$ or $-y_b$, respectively), 
and the other end is fluctuating from source to source in the range $[-y_b,y_b]$. We assume a uniform production of entropy from a source with a random end-point $y$, i.e., 
\begin{eqnarray}
&& f_A(\eta;y)=\theta(y<\eta<y_b), \nonumber  \\
&& f_B(\eta;y)=\theta(-y_b<\eta<y),  \label{eq:ety}
\end{eqnarray}
where $\theta$ is a step function equal 1 when the argument is true, and 0 otherwise.
The random end $y$ is generated according to a suitably chosen distribution $g(y)$.
Averaging over events involves averaging over $y$, therefore
\begin{eqnarray}
&& {\br{f_{A}(\eta)}} \!=\! \int_{-y_b}^{y_b} \!\!\!\!\!\!\! dy \,g(y)  f_{A} (\eta;y) \!=\! \int_{-y_b}^{\eta} \!\!\!\!\!\!\! dy \, g(y) \!=\!G(\eta)-G(-y_b), \nonumber \\
&& {\br{f_{B}(\eta)}} \!=\! \int_{-y_b}^{y_b} \!\!\!\!\!\!\! dy \,g(y)  f_{B} (\eta;y) \!=\! \int_{\eta}^{y_b}  \!\!\!\!\!\!\! dy \,g(y) \!=\! G(y_b)-G(\eta), \nonumber \\ \label{eq:one}
\end{eqnarray}
with $G'(y)=g(y)$. Therefore, to match to Eq.~(\ref{eq:tri}) we must simply take for the length distribution function 
\begin{eqnarray}
g(y)=d/dy {\br{f_{A}(y)}} = - d/dy {\br{f_{B}(y)}}.
\end{eqnarray}

Since in the experimental coverage of the ATLAS experiment \cite{ATLAS:anm} $h(\eta)$ is flat within the range of Eq.~(\ref{eq:flat}) (we may take $h(\eta)=1$ as normalization is 
not relevant), we find
\begin{eqnarray}
g(y)=\frac{1}{2y_b}, \label{eq:enduni}
\end{eqnarray}
i..e., a uniform distribution of the end point, valid in the central pseudorapidity region.

\begin{figure}
\begin{center}
{\includegraphics[width=0.4\textwidth]{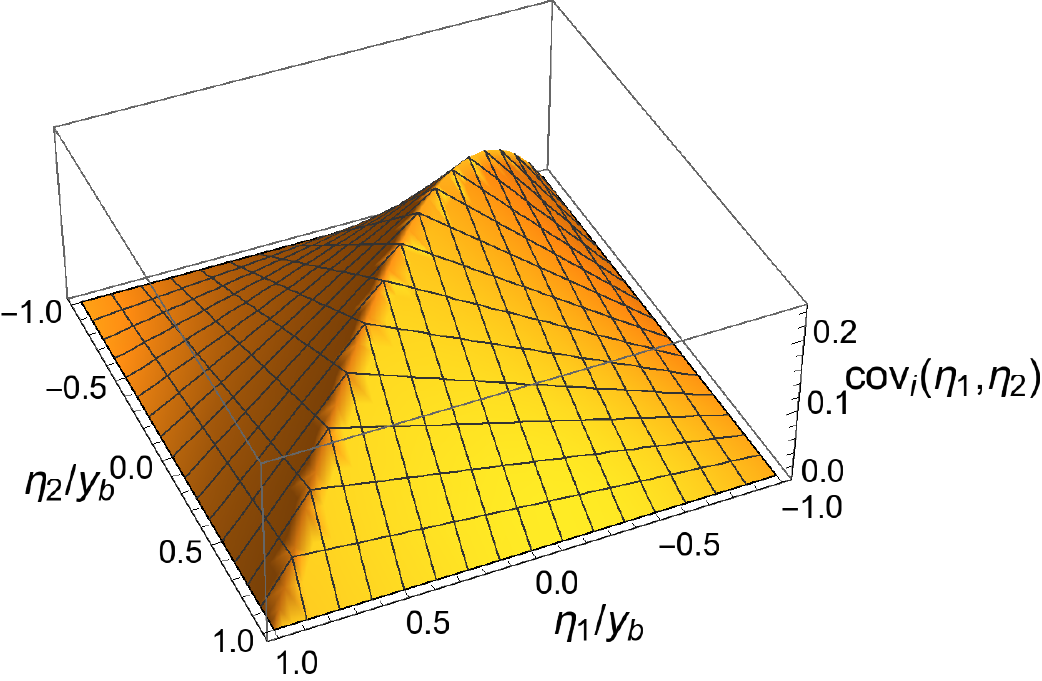}}
\end{center}
\vspace{-5mm}
\caption{(Color online) The covariance from single-end fluctuations, Eq.~(\ref{eq:pcov}).
\label{fig:pc1}}
\end{figure}

The length fluctuations induce correlations, as we have
\begin{eqnarray}
&& \br{f_{A}(\eta_1,\eta_2)} = \int_{-y_b}^{y_b} \!\!\!\!\!\!\! dy \,g(y) f_{A} (\eta_1;y) f_{A}(\eta_2;y) \nonumber \\
&& = \int_{-y_b}^{{\rm min}(\eta_1,\eta_2)} \hspace{-12mm} dy \,g(y) = {\br{f_{A}({\rm min}(\eta_1,\eta_2))}}, \nonumber \\
&& \br{f_{B}(\eta_1,\eta_2)} = \int_{-y_b}^{y_b} \!\!\!\!\!\!\! dy \,g(y) f_{B} (\eta_1;y) f_{B}(\eta_2;y) \nonumber \\
&& = \int_{{\rm max}(\eta_1,\eta_2)}^{y_b} \hspace{-12mm} dy \,g(y) = {\br{f_{B}({\rm max}(\eta_1,\eta_2))}}.
\label{eq:two} 
\end{eqnarray}
In the central region with Eq.~(\ref{eq:enduni}) we have 
\begin{eqnarray}
\br{f_{A}(\eta_1,\eta_2)} &=& t[{\rm min}(\eta_1,\eta_2)],  \nonumber \\ 
\br{f_{B}(\eta_1,\eta_2)} &=& t[-{\rm max}(\eta_1,\eta_2)]. 
\end{eqnarray}
Consequently, we find
\begin{eqnarray}
&& {\rm cov}_{A} (\eta_1,\eta_2) = {\rm cov}_{B} (\eta_1,\eta_2) =  \frac{y_b^2 - {\eta_1 \eta_2} - y_b {|\eta_1-\eta_2|} }{4 y_b^2}, \nonumber \\ \label{eq:pcov}
\end{eqnarray}
where we have used  \mbox{${\rm max}(a,b)=(a+b+|a-b|)/2$} and \mbox{${\rm min}(a,b)=(a+b-|a-b|)/2$}.
Formula (\ref{eq:pcov}) is illustrated in Fig.~\ref{fig:pc1}.

\begin{figure}
\begin{center}
{\includegraphics[width=0.4\textwidth]{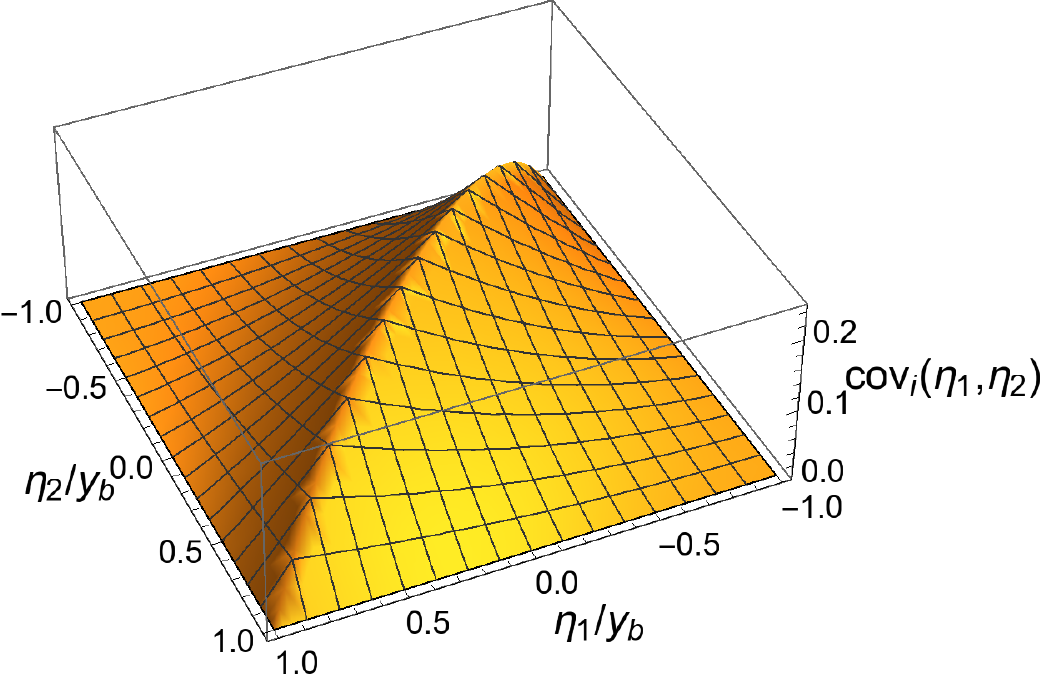}}
\end{center}
\vspace{-5mm}
\caption{(Color online) The covariance from double-end fluctuations, Eq.~(\ref{eq:pcov2}).
\label{fig:pc2}}
\end{figure}

Note that we have obtained correlation, although the emission was independent in the sense that we have used the product $ f_{i} (\eta_1;y) f_{i}(\eta_2;y) $ in 
the integrand in Eq.~(\ref{eq:two}). The correlation is due to the common limit on emission of both particles from the fluctuating end-point $y$.

We also remark that the assumed formula (\ref{eq:tria}) simultaneously fixes the symmetric and antisymmetric parts in Eq.~(\ref{eq:symasym}). In a more general situation 
this need not be the case, for instance we might have a larger symmetric component, replacing 
\begin{eqnarray}
t(\eta;y_b) \to  t(\eta;y_b) + \alpha,
\end{eqnarray}
where $\alpha > 0$. The presence of an extra symmetric component would reduce the correlations coming from the fluctuating end-points, since
$f_s(\eta_1) f_s(\eta_2)$ appears in the denominator of $C(\eta_1,\eta_2)$.

\subsection{Double-end fluctuations \label{sec:double}}

The derivation of the previous section may be straightforwardly extended to the case where both ends fluctuate. Then
\begin{eqnarray}
&& f_A(\eta;y_1,y_2)=\theta(y_1<\eta<y_2), \label{eq:ety2}
\end{eqnarray}
and 
\begin{eqnarray}
&& {\br{f_{A}(\eta)}} = \int_{-y_b}^{y_b} \!\!\!\!\!\!\! dy_1 \,g_1(y_1)  \int_{-y_b}^{y_b} \!\!\!\!\!\!\! dy_2 \,g_2(y_2) f_{A} (\eta;y_1,y_2) \nonumber \\ 
&& = \int_{-y_b}^{\eta} \!\!\!\!\!\!\! dy_1 \, g_1(y_1)   \int^{y_b}_{\eta} \!\!\!\!\!\!\! dy_2 \, g_2(y_2) \nonumber \\
&& = \left [ G_1(\eta)-G_1(-y_b) \right ] \left [ G_2(y_b)-G_2(\eta) \right] \label{eq:one2}
\end{eqnarray}
(similar expressions hold for $\br{f_{B}(\eta)}$). In the present case, in general, we cannot uniquely obtain $G_1(y)$ and  $G_2(y)$ by
matching to $\br{f_{A}(y)}$. However, in the case where one end of the source is close to the fragmentation region (as  expected of wounded quarks, for example), its fluctuations do not 
enter the central region (\ref{eq:flat}) and effectively, for that domain, we get the model with single-end fluctuations of Sec.~\ref{sec:single}.

In the special case where both end-points fluctuate uniformly, we get from Eq,~(\ref{eq:one2}) the symmetric component 
\begin{eqnarray}
f_s(\eta)=\frac{y_b^2-\eta^2}{4 y_b^2},
\end{eqnarray}
as well as the contribution to the covariance 
\begin{eqnarray}
{\rm cov}_i(\eta_1,\eta_2) =  \frac{y_b^2-{\rm min}(\eta_1,\eta_2)^2}{4 y_b^2} - \frac{y_b^2-\eta_1^2}{4 y_b^2}\frac{y_b^2-\eta_2^2}{4 y_b^2}. \nonumber \\
\label{eq:pcov2}
\end{eqnarray}
Equation  (\ref{eq:pcov2}) is illustrated in Fig.~\ref{fig:pc2}. We note that the shape of the obtained covariance is qualitatively similar to the case of Fig.~\ref{fig:pc1}, with the size smaller by 25\%, and 
a faster fall-off in $|\eta_1-\eta_2|$ for the case of double-end fluctuations.

\section{Fluctuating strength \label{sec:strength}}

In Glauber-like models, a common ingredient is the {\em overlaid} distribution, i.e., on top of the sources 
we superpose a random distribution~\cite{Broniowski:2007nz,*Rybczynski:2013yba}. This describes the possibility that the sources may have randomly varying strength. 
The superposition is necessary to properly describe the multiplicity distributions of produced hadrons when the conventional Glauber model is used~\cite{Bozek:2013uha}.

In our framework, for the model with single-end fluctuations, the fluctuating strength is incorporated as a generalization of Eq.~(\ref{eq:ety}), namely
\begin{eqnarray}
&& f_A(\eta;y)={\omega} \theta(y<\eta<y_b), \nonumber \\
&& f_B(\eta;y)={\omega} \theta(-y_b<\eta<-y),
\end{eqnarray}
where $\omega$ is a random variable giving the strength of the source. 
With these extra fluctuations we have the following generalization of Eqs.~(\ref{eq:one},\ref{eq:pcov}):
\begin{eqnarray}
&& {\br{f_{A,B}(\eta)}} = \br{\omega} t(\pm \eta;y_b), \nonumber \\
&& {\rm cov}_{A,B} (\eta_1,\eta_2) = \br{\omega^2} \frac{y_b^2 - {\eta_1 \eta_2} - y_b |\eta_1-\eta_2| }{4 y_b^2} \nonumber \\ && 
+{\rm var}(\omega)  \frac{y_b^2 +{\eta_1 \eta_2} \pm  y_b (\eta_1+\eta_2) }{4 y_b^2}. \nonumber  \\
\end{eqnarray}

The formulas for the correlation functions are given in Eq.~(\ref{eq:gen2}-\ref{eq:genasym2}). For the marginal projections of Eq.~(\ref{eq:cp}) 
we find for the symmetric case
\begin{eqnarray}
C_p(\eta)&=&1+\frac{{\rm var}(N_+)}{\br{N_+}^2}+\frac{s(\omega)}{\br{N_+}}\nonumber \\ &+& r [1+s(\omega)] \frac{(2y_b-Y)Y-\eta^2}{2\br{N_+}y_b Y}, \label{eq:cpsym}
\end{eqnarray}
and for the asymmetric case
\begin{eqnarray}
&& C_p(\eta)=1+\frac{{\rm var}(N_A)}{\br{N_A}^2}+\frac{s(\omega)}{\br{N_A}} \label{eq:cpasym} \\ &+& 
r [1+s(\omega)] \frac{\eta(y_b-Y)+y_b(\eta+y_b)\log \left( \frac{Y+y_b}{\eta+y_b} \right )}{\br{N_A}(\eta+y_b) Y}, \nonumber
\end{eqnarray}
where $r$ is defined above Eq.~(\ref{eq:u}).

\section{Gaussian smearing}

The application of the Gaussian smearing procedure of Eq.~(\ref{eq:smear}) is very simple for the considered model. In the numerator of $C(\eta_1,\eta_2)$ it amounts to 
replacing
\begin{eqnarray}
|\eta_1-\eta_2| \to  \frac{2 \sigma_\eta e^{-\frac{(\eta_1-\eta_2)^2}{4 \sigma_\eta^2}}}{\sqrt{\pi }}+
(\eta_1-\eta_2) \, {\rm erf}\left(\frac{\eta_1-\eta_2}{2  \sigma_\eta}\right), \nonumber \\
\end{eqnarray}
with the other terms ($\eta_1$, $\eta_2$, $\eta_1 \eta_2$) unchanged.

\bibliography{hydr}

\end{document}